\documentclass[conference]{IEEEtran}

\usepackage{cite}
\usepackage{amsmath,amssymb,amsfonts}
\usepackage{algorithmic}
\usepackage{graphicx}
\usepackage{textcomp}
\usepackage{xcolor}
\usepackage{caption}

\begin{document}

\title{Cloud Computing --- Everything As A Service\\
}

\author{\IEEEauthorblockN{Michael Howard}
\IEEEauthorblockA{\textit{Computer Science Department} \\
\textit{Portland State University}\\
Portland, USA \\
mihoward@pdx.edu}
}

\maketitle

\begin{abstract}\label{sec:abs}
Compute infrastructure hosted by a cloud provider allows an application to scale without limit.  The application developer no longer needs to worry about the up-front investment in a server farm provisioned for a worst-case load scenario.  However, managing cloud deployments requires a sophisticated framework that can autoscale the infrastructure and guarantee the up-time of running container images.  This paper surveys existing research addressing the management and orchestration of cloud deployments as well as the modelling framework to abstract away the low-level details of the host infrastructure.  We investigate blockchain distributed ledgers, quantum computing and Internet of Things application stacks to show how they can utilize cloud deployments.
\end{abstract}


\section{Introduction}\label{sec:intro}
Cloud computing enables a new method to provision infrastructure and deploy software.  Compute resources are now commodities that expand to scale with the software application requirements.  These resources temporarily launch on-demand from the cloud host and provide the convenience of not needing to pre-purchase and maintain on-premise infrastructure.

However, since the cloud infrastructure is provisioned and destroyed frequently, there are additional requirements for system management and security.  An ideal approach fully automates the cloud usage, a new application builds and deploys easily while immediately launching compute resources to meet its needs.  This paper surveys existing research on the problem of deploying an application into the cloud environment and adapting to its changing system needs.  

Section \ref{sec:tee} is reviewing the Continuous Integration/Continuous Deployment (CI/CD) pipeline and the Tekton framework to drive it.  The container images deployed into the cloud can also enhance their runtime security through a Trusted Execution Environment (TEE). Section \ref{sec:cloudcamp} introduces the CloudCAMP framework that provides a graphical modelling environment abstracting the provisioning and deployment of cloud infrastructure.  After provisioning the cloud infrastructure, the orchestration cluster manages all running containers.  Section \ref{sec:orchestration} compares the top orchestration frameworks: Kubernetes and Docker Swarm.

Next, the survey moves on to how different application types can manage cloud deployments.  In Section \ref{sec:blockchain}, the application is a blockchain distributed ledger recording new transactions coming from Renault cars equipped with Internet of Things (IoT) devices sending sensor data from a collision.  Section \ref{sec:quantum} discusses a paper focused on the unique needs of quantum-compute applications in the cloud.  The authors' extend an existing orchestration specification to create Topology and Orchestration Specification for Cloud Applications for Quantum Computing (TOSCA4QC).  Multiple TOSCA4QC case studies prove the practical feasibility of this new specification.  Finally, Section \ref{sec:sis} addresses the orchestration of smart IoT systems with a paper on a framework called Generation and Deployment of Smart IoT Systems (GeneSIS).  It uses a modelling language to generate the modules needed for the deployment and dynamic adaption of heterogeneous IoT systems.   

\section{Trusted Execution Environment}\label{sec:tee}
Most cloud providers handle deploying an application into a cloud environment in an automated fashion.  However, the resulting vendor lock-in is counter-productive to most application developers.  As a result, the authors of \emph{``A Kubernetes CI/CD Pipeline with Asylo as a Trusted Execution Environment Abstraction Framework"} \cite{mahboob2021} discuss the benefits of using a third-party service --- Tekton --- to configure the triggers and actions of a Continuous Integration/Continuous Deployment (CI/CD) pipeline.  They also discuss building the runtime containers with the Asylo framework.  

Tekton is a framework that is publicly exposed through a hosted service.  It provides a declarative language to define each stage of the CI/CD pipeline.  These stages run as containerized tasks within a Kubernetes cluster and each task listens for its corresponding trigger events.  The task executes the appropriate pipeline run action upon receiving the trigger event.  The task containers perform the following actions:
\begin{enumerate}
    \item Checkout the source code from source-control management.
    \item Build the source code binaries.
    \item Execute the unit test suite.
    \item Build the execution container with Kaniko and using Asylo as a binary wrapper. \label{itm:asylo}
    \item Archive the execution container in the Harbor registry.
    \item Deploy the execution container into a runtime environment for functional testing.
    \item Promote (deploy) the execution container into other runtime environments for end-user interaction.
\end{enumerate}

A significant benefit to using Tekton is avoiding locking in to a specific cloud vendor.  Amazon Web Services (AWS) has its Code Pipeline service and Google Cloud Platform provides Cloud Build; both are excellent tools for implementing a CI/CD pipeline.  However, switching between cloud vendors requires porting all the pipeline configuration and code.  The Tektron framework is vendor-agnostic and easily switches between cloud deployment targets. 

The Asylo framework, mentioned in Item \ref{itm:asylo} above, provides a wrapper around the application binaries such that they run within a Trusted Execution Environment (TEE).  Thus, binaries running within the container image utilize Intel's Security Guard Extension (SGX) to encrypt instructions as they page into memory and then decrypt them as they load in their corresponding processor.  This memory-encryption feature prevents a cloud provider's system administrator from being able to see or interfere with instructions once they are loaded into Virtual Machine memory. 

\begin{figure}[htbp]
    \centerline{\includegraphics[width=.9\linewidth, keepaspectratio]{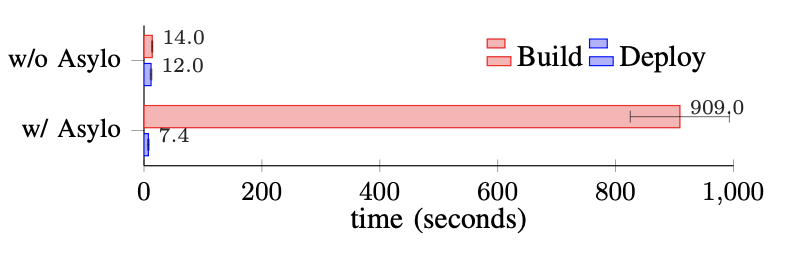}}
    \captionsetup{width=.8\linewidth}
    \caption{Build and deploy times for a pipeline experiment as discussed in Mahboob \cite{mahboob2021}.  The experiment builds a container image with and without Asylo and benchmarks the execution time.}
    \label{fig:asylo}
\end{figure}

The Asylo wrapper builds into the execution container and allows the use of SGX instructions, even if the application is not compiled with the SGX option. Additionally, there is an emulation flag allowing deployment to Virtual Machines that do not support SGX.  This emulation is useful when testing the execution container in different environments.  Figure \ref{fig:asylo} shows the performance cost incurred with Asylo during the benchmark experiments.  Deployment time of the execution container image does not change significantly when including Asylo; however, the build times did increase by a factor of 65$\times$.  Thus, there is a trade-off between slightly increasing development time and a more secure runtime. 

\section{CloudCAMP Provisioning}\label{sec:cloudcamp}
CloudCAMP is another framework tool, proposed at Vanderbilt University, that focuses on the management of cloud deployments.  The paper \emph{``CloudCAMP: Automating Cloud Services Deployment \& Management"} \cite{bhattacharjee2019} goes into detail on the authors' vision of a technology-agnostic framework that avoids relying on domain expertise of the target cloud infrastructure.

The framework works on a model-driven approach that generates Infrastructure as Code (IaC) modules for provisioning cloud services.  A Graphical User Interface (GUI) allows editing the model and provides a high level of abstraction without requiring low-level scripting or cloud vendor domain knowledge.  

\begin{figure}[htbp]
    \centerline{\includegraphics[width=.9\linewidth, keepaspectratio]{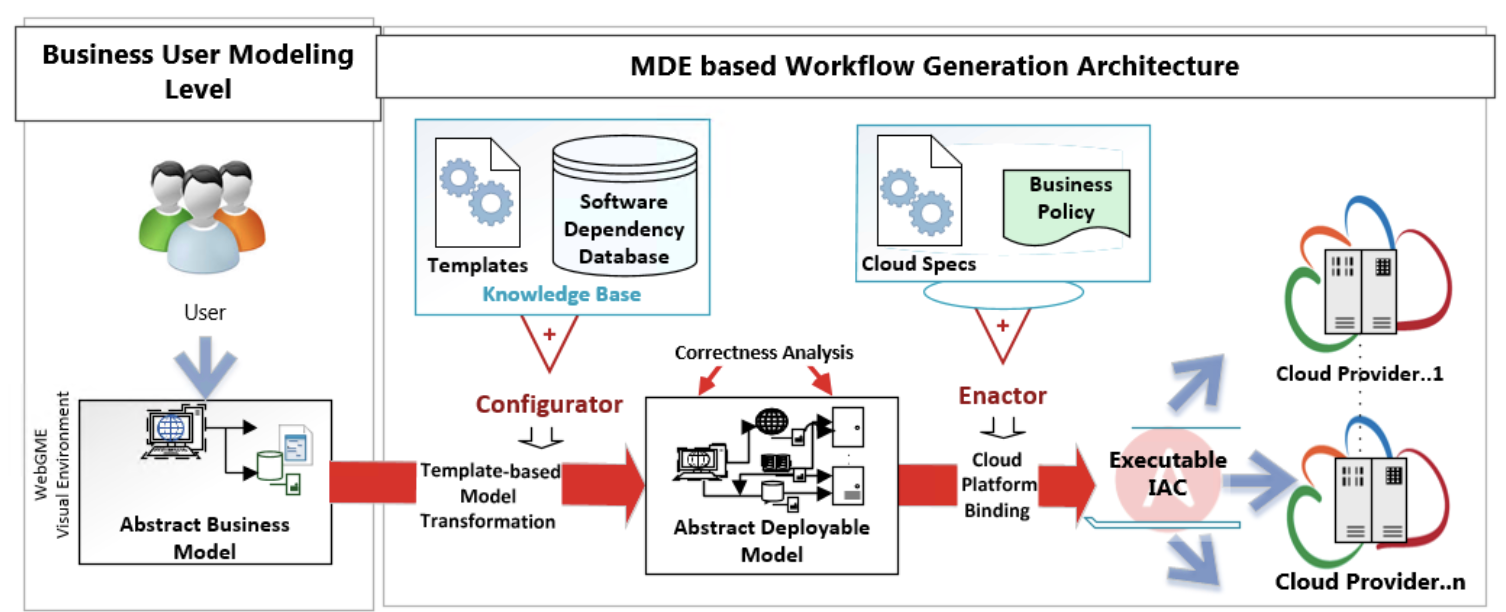}}
    \captionsetup{width=.8\linewidth}
    \caption{The CloudCAMP workflow, shown in Bhattacharjee \cite{bhattacharjee2019}, describes a typical sequence starting from the user input to the generation of provider-specific IaC modules.}
    \label{fig:cloudcamp}
\end{figure}

CloudCAMP differs from Tekton by focusing on the modelling and deployment of the cloud infrastructure and application stack.  Tekton implements the pipeline stages for CI/CD.  It can also potentially trigger cloud infrastructure provisioning via pre-configured CloudCAMP modules (shown in Figure \ref{fig:cloudcamp}) that would then host the deployed pipeline container images.

The building blocks in the modelling environment are all the application components; each connect with exposed endpoints.  These components form the basis of the Domain Specific Modelling Language (DSML) that drives the CloudCAMP framework. The WebGME user interface provides a way to edit DSML.  It is a cloud-based framework for DSML developers that defines their language and create model parsers for code artifact generators.  Each building block in the model forms a node.  Many node types are supported such web, database and analytics applications.  Multiple cloud providers for each type are also supported such as OpenStack, Amazon Web Services and Microsoft Azure. 

\section{Container Orchestration Tools}\label{sec:orchestration}
After the CI/CD pipeline deploys the container images, they additionally require orchestration on the provisioned cloud infrastructure.  The orchestration monitors as well as manages the application and system requirements of each running container image.  \emph{``A Performance Comparison of Cloud-based Container Orchestration Tools''} \cite{pan2019} provides a review and comparison of Kubernetes versus Docker Swarm, two of the top orchestration tools currently in use.  The authors conclude through their benchmark tests that Kubernetes is the more scalable orchestrator.  However, it can introduce as much as 8.3\% performance penalty on small-scale systems compared with Docker alone (no orchestration).

\begin{figure}[htbp]
    \centerline{\includegraphics[width=.9\linewidth, keepaspectratio]{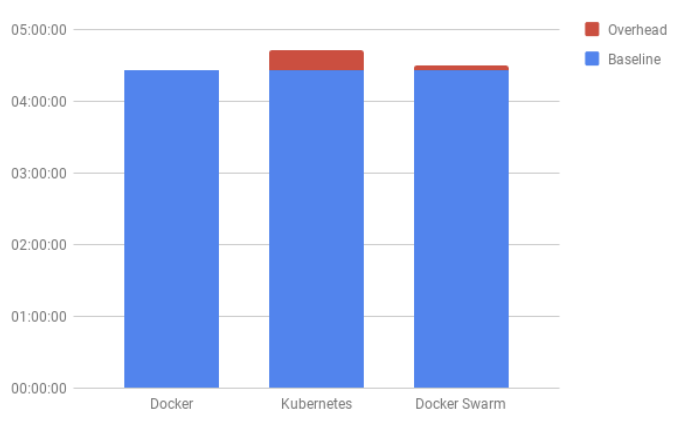}}
    \captionsetup{width=.8\linewidth}
    \caption{Orchestrator overhead as shown in Pan \cite{pan2019}.  The graph shows the processing time of a sample containerized application that is deployed directly, then again with Docker Swarm and Kubernetes orchestration.}
    \label{fig:overhead}
\end{figure}

An application spanning multiple containers running over a cluster of cloud-based worker nodes evaluates and benchmarks the performance of the container orchestrators.  First, running the container images directly on the worker nodes without any orchestration determines a baseline.  A benchmarking test suite captures processing times and system utilization metrics.  The two applications in the benchmarking tests are:
\begin{enumerate}
    \item A Phoronix Test Suite (details in Pan \cite{pan2019}).
    \item A Light Detection and Ranging (LiDAR) data processing.
\end{enumerate}
    
The results in Figure \ref{fig:overhead} show processing times with container images running directly on the node and again orchestrated through Docker Swarm and Kubernetes.  The red portion indicates the overhead processing that the orchestrator introduces and shows Kubernetes with the highest overhead.

\section{Blockchain Distributed Ledger}\label{sec:blockchain}
So far, the focus has been on the technology of the development pipeline and the provisioning of the hosting cloud environment.  \emph{``A Blockchain cloud architecture deployment for an industrial IoT use case''} \cite{gerrits2021} discusses how a specific application is using that technology.  The authors study the feasibility of using a cloud Kubernetes environment to host a blockchain-based, distributed ledger called Hyperledger Sawtooth.

\begin{figure}[htbp]
    \centerline{\includegraphics[width=.9\linewidth, keepaspectratio]{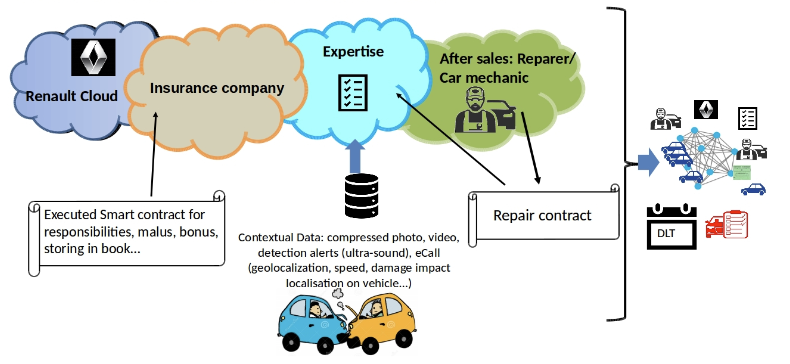}}
    \captionsetup{width=.8\linewidth}
    \caption{Hyperledger Sawtooth uses blockchains for automatically declaring accidents in Renault's vehicular infrastructure as shown in Gerrits \cite{gerrits2021}.}
    \label{fig:renault}
\end{figure}

The use case for this study involves Renault cars containing Internet of Things (IoT) devices able to connect to a cloud-hosted blockchain.  Each car sends sensor telemetry data before and after a collision using a smart contract.  The smart contract enables adding a new blockchain transaction to the ledger.  Renault, insurance companies, mechanics and police can all access the ledger to determine which car is at fault in the collision.  Figure \ref{fig:renault} shows the ecosystem of clients connecting to the ledger.

The Kubernetes cluster used in the study contained six cloud instances.  Within that cluster, blockchain Sawtooth nodes run independently, each within their own Docker container image.  The number of nodes varies between 4 and 24, while the commit rate of transactions increased from 5 to 50 per second.  The study concluded that the Sawtooth core itself was the bottleneck and a newer version fully coded in Rust and supporting multithreading would provide faster blockchain commit rates.  The cloud infrastructure in the study also proved to successfully handle the distributed ledger.

\section{Quantum Computing}\label{sec:quantum}
Another application that can take advantage of cloud deployment is quantum computing.  The paper \emph{``TOSCA4QC: Two Modeling Styles for TOSCA to Automate the Deployment and Orchestration of Quantum Applications''} \cite{wild2020} starts by discussing a technology called Topology and Orchestration Specification for Cloud Applications (TOSCA) that takes a declarative approach to deploying classical (non-quantum) applications.

\begin{figure}[htbp]
    \centerline{\includegraphics[width=.9\linewidth, keepaspectratio]{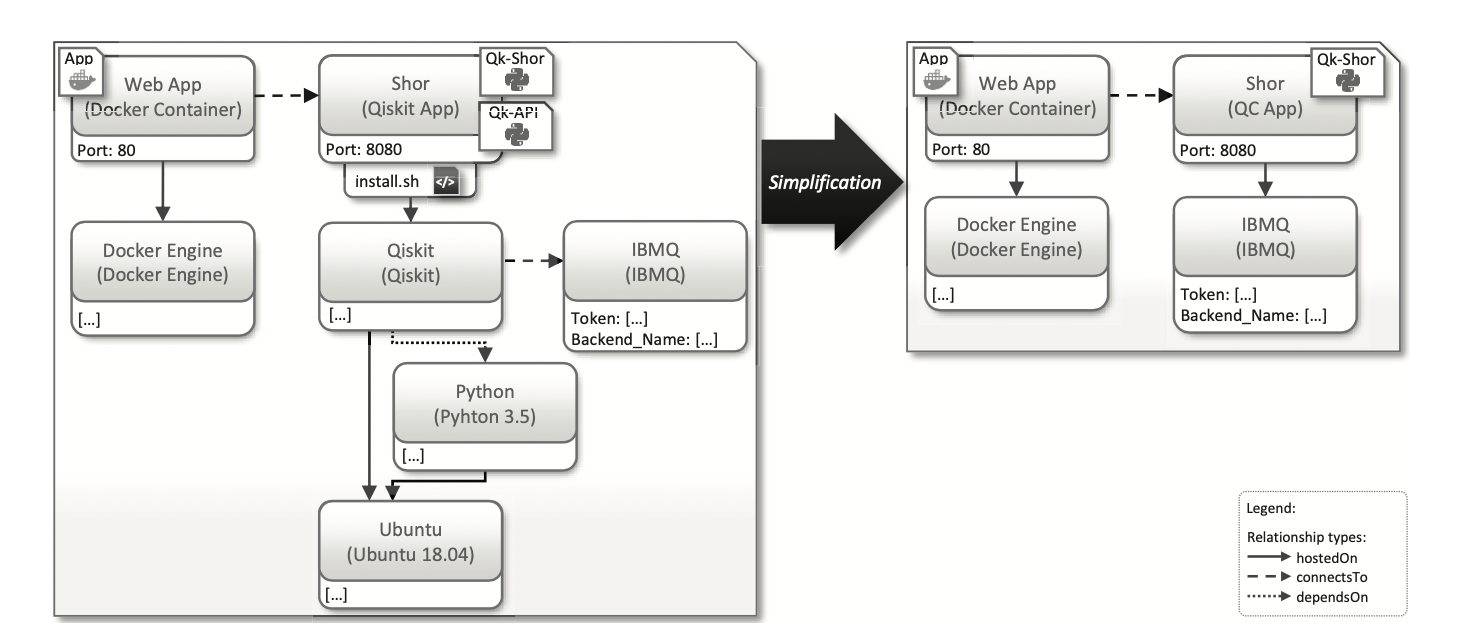}}
    \captionsetup{width=.8\linewidth}
    \caption{TOSCA4QC SDK-specific (SDK-S) and SDK-agnostic (SDK-A) modelling styles as shown in Wild \cite{wild2020}.}
    \label{fig:tosca4qc}
\end{figure}

Quantum applications must deploy for each invocation.  This restriction means deploying only once and not invoking multiple times per deployment, such as with classical application deployments.  To meet this challenge, the authors extend TOSCA to introduce Topology and Orchestration Specification for Cloud Applications for Quantum Computing (TOSCA4QC).  The new extended standard introduces two deployment modelling styles:
\begin{enumerate}
    \item The Software Development Kit (SDK)-specific, which covers all technical details.
    \item The SDK-agnostic, which reflects common modelling principles known from classical applications while hiding technical details.
\end{enumerate}
\noindent Figure \ref{fig:tosca4qc} shows the GUI and abstract modelling using SDK-specific style on the left and the SDK-agnostic on the right.

The Software Development Kit Agnostic (SDK-A) style does not provide enough detail to make the model executable (able to deploy).  It follows the common modelling approach of a classical application.  To make it executable, it first transforms into a Software Development Kit Specific (SDK-S) model.  The TOSCA4QC standard provides an automated approach to transform the SDK-A model to SDK-S.  Thus, the standard allows modelling of quantum cloud deployments in a simple and easy to understand format then another, more complicated format that is also executable.

\section{Smart IoT Systems}\label{sec:sis}
Multiple tools have emerged to support the continuous deployment of cloud-based software systems.  However, there is currently a lack of proper tool support for the continuous orchestration and deployment of software systems in the IoT space.  \emph{``GeneSIS: Continuous Orchestration and Deployment of Smart IoT Systems''} \cite{ferry2019} addresses these IoT systems that may have little or no direct access to the Internet.

The authors of this paper developed a framework called Generation and Deployment of Smart IoT Systems (GeneSIS) that allows decentralized processing across heterogeneous IoT systems.  The GeneSIS framework includes:
\begin{enumerate}
    \item A domain-specific modelling language to model the orchestration and deployment of a Smart IoT System.
    \item The execution engine to drive that orchestration.
\end{enumerate}

\begin{figure}[htbp]
    \centerline{\includegraphics[width=.9\linewidth, keepaspectratio]{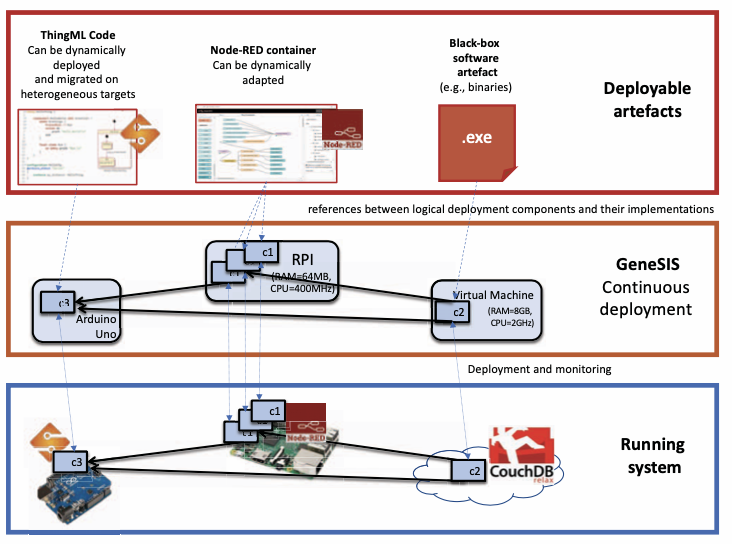}}
    \captionsetup{width=.8\linewidth}
    \caption{Overall GeneSIS approach as shown in Ferry \cite{ferry2019}.  It summarizes modelling the IoT targets and deploying their software artifacts.}
    \label{fig:genesis}
\end{figure}

\noindent Figure \ref{fig:genesis} shows the authors' vision of how the GeneSIS modelling language is sufficient to support the deployment and dynamic adaption of IoT systems.

This paper presents how GeneSIS leverages model-driven techniques and methods to support the continuous deployment and orchestration of Smart IoT Systems (SIS).  The modelling language supports the deployment of SIS over IoT, edge and cloud infrastructure in a platform-independent way.  The execution engine provides the mechanism to execute this deployment and support its dynamic adaption, even if the device has limited internet access. 

\section{Conclusion and Discussion}
The papers in this survey show a new paradigm in software development using cloud computing for deployment.  Purchasing cloud compute resources allows the development and deployment to begin with only provisioning the minimum resources needed for the best-case system load.  Thus, it is cost-effective compared with building up an on-premise compute farm that requires up-front budgeting and provisioning to handle the worst-case system load.  Another benefit is the cloud does not require maintenance as with the compute farm.  For instance, operating system updates incur down-time, which is avoided in the cloud if purchasing container orchestration services rather than the virtual machines directly.

The cloud-compute resources are dynamic and therefore provisioned and destroyed frequently.  This on-demand provisioning requires additional management software to control the system configuration and security.  The CloudCAMP framework allows modelling the cloud infrastructure at a high level, agnostic of the cloud provider.  A container wraps the software application  and encapsulates the runtime environment.  A Continuous Integration/Continuous Deployment pipeline automates the building and deployment of the application containers.  Tekton implements the pipeline while avoiding locking in to a single cloud vendor.  A Trusted Execution Environment enhances runtime security, implemented by wrapping the container's application binaries with the Asylo framework.  Kubernetes orchestrates the container deployment while introducing a runtime overhead due to its flexible feature set.

In addition to how cloud computing is implemented, there are also three application stacks included in the survey.  Quantum computing takes advantage of cloud resources through a new standard called Topology and Orchestration Specification for Cloud Applications for Quantum Computing.  Internet of Things devices use a framework called Generation and Deployment of Smart IoT Systems which focuses on limited or no internet access scenarios.  Finally, Renault cars use a blockchain-based Hyperledger Sawtooth distributed ledger to record collision event data.

Future research needs to focus on splitting existing applications into discrete microservices such that they may be easily containerized.  Other areas to include are implementing smart orchestration such that deployed containers run more robustly through redundant execution and intelligent routing of network requests to assist with load balancing.  Machine Learning provides an excellent framework to train the orchestrator on optimizing the operating parameters and is a logic next step in the expansion of cloud computing. Everything as a Service is within reach.

\section{Acknowledgements}\label{sec:ack}
The author would like to acknowledge the contributions of Dr. Wu-Chang Feng at Portland State University.  His advisorship, funding and support made this research possible.

\vspace{12pt}

\end{document}